\documentclass[longbibliography,superscriptaddress,twocolumn]{revtex4-2}
\usepackage[utf8]{inputenc}

\usepackage{amsmath}
\usepackage{amssymb}
\usepackage{mathtools}
\usepackage[colorlinks]{hyperref}
\usepackage{soul}

\DeclarePairedDelimiter\abs{\lvert}{\rvert}
\DeclarePairedDelimiter\set{\{}{\}}
\DeclarePairedDelimiter\parens{(}{)}
\DeclarePairedDelimiter\bracks{[}{]}

\DeclarePairedDelimiter\ket{\lvert}{\rangle}

\DeclarePairedDelimiter\ceil{\lceil}{\rceil}
\DeclareMathOperator{\tr}{Tr}

\newcommand{\eps}{\varepsilon}
\newcommand{\ph}{\varphi}
\newcommand{\vac}{\ket{\text{vac}}}

\newcommand{\calB}{\ensuremath{\mathcal{B}}}

\newcommand{\calG}{\ensuremath{\mathcal{G}}}

\newcommand{\calV}{\ensuremath{\mathcal{V}}}
\newcommand{\calU}{\ensuremath{\mathcal{U}}}
\newcommand{\calW}{\ensuremath{\mathcal{W}}}

\begin{document}

\title{Simulating Majorana zero modes on a noisy quantum processor}
\author{Kevin J. Sung}
\email[email: ]{kevinsung@ibm.com}
\affiliation{IBM Quantum, IBM T.J. Watson Research Center, Yorktown Heights, NY 10598, USA}
\author{Marko J. Ran\v{c}i\'{c}}
\affiliation{TotalEnergies, Tour Coupole La Défense, 2 Pl. Jean Millier, 92078 Paris, France}
\author{Olivia T. Lanes}
\affiliation{IBM Quantum, IBM T.J. Watson Research Center, Yorktown Heights, NY 10598, USA}
\author{Nicholas T. Bronn}
\affiliation{IBM Quantum, IBM T.J. Watson Research Center, Yorktown Heights, NY 10598, USA}

\date{\today}

\begin{abstract}
    The simulation of systems of interacting fermions is one of the most anticipated
    applications of quantum computers. The most interesting simulations will require
    a fault-tolerant quantum computer, and building such a device remains a long-term
    goal. However, the capabilities of existing noisy quantum processors have steadily
    improved, sparking an interest in running simulations that, while not necessarily
    classically intractable, may serve as device benchmarks and help elucidate the
    challenges to achieving practical applications on near-term devices.
    Systems of \emph{non}-interacting fermions are ideally suited to serve these
    purposes. While they display rich physics and generate highly entangled states
    when simulated on a quantum processor, their classical tractability enables
    experimental results to be verified even at large system sizes that would
    typically defy classical simulation. In this work, we use a noisy
    superconducting quantum
    processor to prepare Majorana zero modes as eigenstates of the Kitaev chain
    Hamiltonian, a model of non-interacting fermions. Our work builds on
    previous experiments with non-interacting fermionic systems. Previous work
    demonstrated error mitigation techniques applicable to the special case of
    Slater determinants. Here, we show how to extend these techniques to the
    case of general fermionic Gaussian states, and demonstrate them by preparing
    Majorana zero modes on systems of up to 7 qubits.
\end{abstract}

\maketitle

\section{Introduction}

\begin{figure*}
\includegraphics[width=1.0\textwidth]{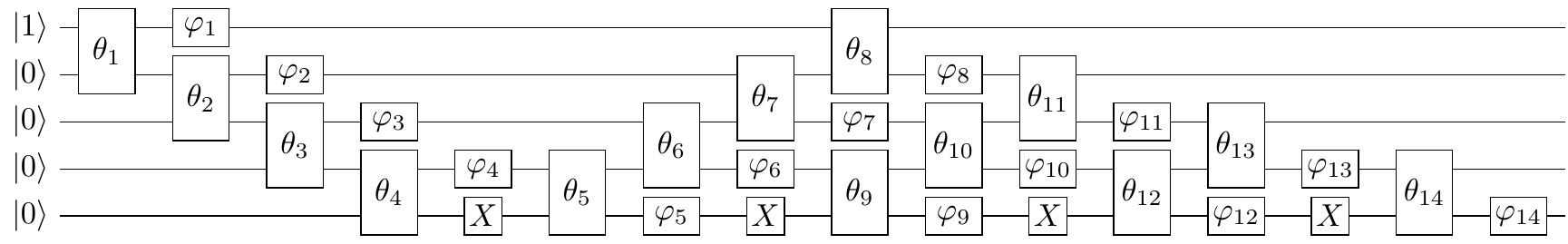}
\caption{Circuit to prepare the first excited state of a 5-mode Kitaev chain.
Gates labeled with an angle $\theta$ are real-valued Givens rotation gates
and gates labeled with an angle $\varphi$
are $Z$ rotations. Note that the presence of the $X$ gates implies that
in general, the circuit does not conserve particle number.
However, the circuit always prepares a state with a well-defined parity.
The Givens rotation gate can be implemented on IBM hardware
by decomposing it into 2 CNOT gates plus single-qubit rotations.
}
\label{fig:circuit}
\end{figure*}

\begin{figure}[b]
\includegraphics[width=1.0\linewidth]{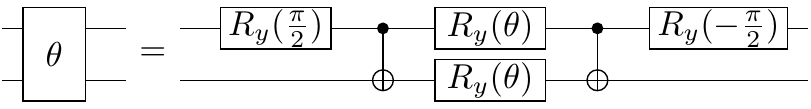}
\caption{Decomposition of the Givens rotation gate into two CNOT gates plus
single-qubit rotations.}
\label{fig:xx_plus_yy}
\end{figure}

The simulation of systems of interacting fermions is one of the most anticipated applications of
quantum computers due to its value to commercial industry and
scientific research~\cite{aspuruguzik2005simulated,georgescu2014quantum}.
The most interesting simulations will undoubtedly require a fault-tolerant quantum
computer capable of executing arbitrarily long quantum programs.
The effort to build such a device is underway
at academic and industrial institutions, and while fault-tolerance remains a long-term goal,
the capabilities of existing prototypes have steadily
improved~\cite{preskill2018quantum,arute2019quantum,wright2019benchmarking,jurcevic2021demonstration,pino2021demonstration}.
These improved capabilities have sparked an interest in running simulations that,
while not necessarily classically intractable,
may serve as device benchmarks and help elucidate
the challenges to achieving practical applications on near-term
devices~\cite{omalley2016scalable,kandala2017hardware,hempel2018quantum,kandala2019error,arute2020hartreefock}.

Systems of \emph{non}-interacting fermions are ideally suited to serve these purposes.
Despite being classically tractable, they display rich physics and
produce highly entangled states when simulated on a quantum processor.
Because they are classically tractable, experimental results can be verified
even at large system sizes that would typically defy classical simulation.
Previous experimental demonstrations of such simulations
include an implementation of the Hartree-Fock method on a
quantum processor~\cite{arute2020hartreefock} and the preparation of Majorana zero modes~\cite{rancic2021exact}.
In both of these experiments, the quantum states prepared and measured belong to
the class of fermionic Gaussian states, of which Slater determinants are a special case.
Fermionic Gaussian states refer to eigenstates of a quadratic Hamiltonian,
the defining feature of a system of non-interacting fermions.

Reference~\cite{arute2020hartreefock} demonstrated
that for Slater determinants, the error mitigation techniques of
physically-motivated postselection of bitstrings and state purification
can be used to significantly improve the fidelity of the simulation and the
accuracy of measured observables. In this work, we show that these
techniques can be extended to the case of general fermionic Gaussian states, and demonstrate them
by improving on the preparation of Majorana zero modes performed in Reference~\cite{rancic2021exact}. We also apply some additional error mitigation
techniques which were not used in either reference.
While Reference~\cite{rancic2021exact} ran experiments on only 3 qubits,
here our error mitigation techniques enable us to go up to 7 qubits
while also obtaining more accurate results.
Our experiments are performed on a superconducting qubit processor
manufactured at IBM.

Majorana zero modes (MZMs) refer to zero-energy Majorana fermion modes that
exhibit topological properties due to
spatial separation of the modes. A prototypical system that contains MZMs is
the Kitaev chain.
The Hamiltonian of a Kitaev chain is
\begin{align}\label{eq:ham}
    \begin{split}
    H = -t &\sum_{j = 1}^n \parens*{a_j^\dagger a_{j + 1} + a_{j + 1}^\dagger a_{j}}\\
    &+ \sum_{j = 1}^n \parens*{\Delta a_j^\dagger a_{j + 1}^\dagger + \Delta^* a_{j + 1} a_j}\\
    &+ \mu \sum_{j = 1}^n \parens*{a_j^\dagger a_j - \frac12},
    \end{split}
\end{align}
where $t$ is the tunneling amplitude, $\Delta$ is the superconducting pairing, $\mu$ is the
chemical potential, and the $\set{a_j}$ are fermionic annihilation operators
for a system of $n$ fermionic modes.
When $\abs{\Delta} = t > 0$ and $\mu = 0$, the Hamiltonian~(\ref{eq:ham}) takes the form
\begin{align}\label{eq:ham_maj}
    H = i t \sum_{j=2}^{2n-2}\gamma_j\gamma_{j+1}
\end{align}
where we have introduced the Majorana fermion operators
\begin{align}
    \gamma_{2j - 1} = a_j + a^\dagger_j, \quad \gamma_{2j} = -i \parens*{a_j - a^\dagger_j}.
\end{align}
Note that $\gamma_1$ and $\gamma_{2n}$ do not appear in the Hamiltonian~(\ref{eq:ham_maj});
these are unpaired zero-energy Majorana modes localized at the ends of the chain.
The energy and separation of the modes
is robust to small perturbations in $\mu$.
MZMs can theoretically be used as carriers of quantum information with a Clifford gate set which is topologically protected
against errors; this fact has motivated efforts at their experimental
realization~\cite{lutchyn2018majorana}.

In this work, we map the Kitaev chain model to a system of qubits
using the Jordan-Wigner transformation (JWT). Despite the nonlocal nature
of the JWT, eigenstates of the model can be prepared efficiently
using only gates acting on neighboring qubits on a line.
We measure the excitation energies of the model by separately preparing
the ground and excited states, and we observe the presence of zero-energy
excitations that are robust to perturbations in $\mu$.
We also measure the Majorana site correlation
$i \gamma_1 \gamma_j$ and observe an exclusive correlation between
the ends of the wire at $\mu=0$ which breaks down with increasing $\mu$.

\section{Results}

\begin{figure*}
\includegraphics[width=0.7\textwidth]{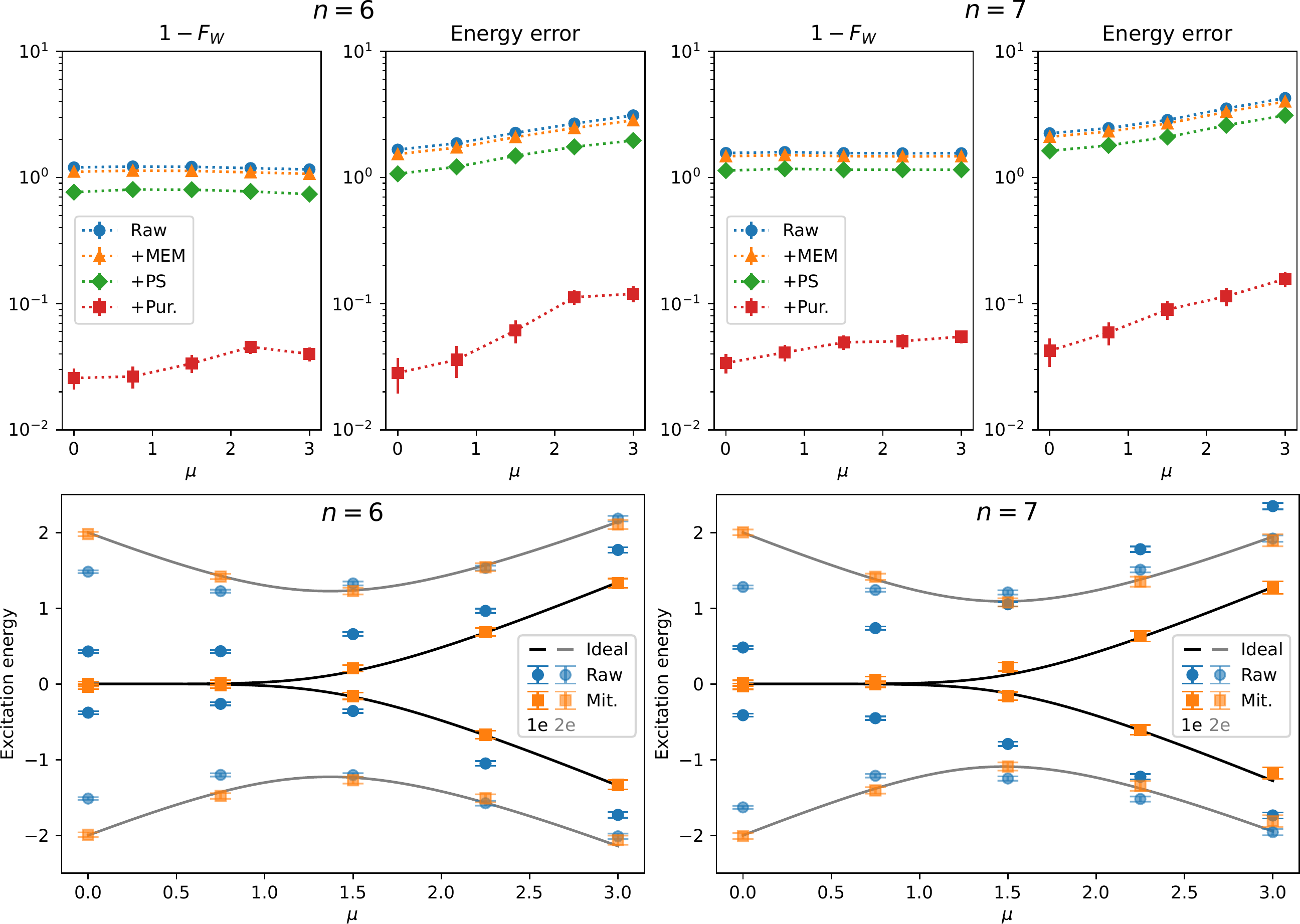}
\caption{Fidelity, error, and energy. (Top) One minus the fidelity witness
and the average error in energy (expectation value of the Hamiltonian).
Note that the fidelity witness is only guaranteed to be a lower bound on
the true fidelity, so it can be negative.
The data for the 6-mode Kitaev chain is shown
on the left, and the data for the 7-mode Kitaev chain is shown on the right.
The different lines in each figure correspond to successive applications
of error mitigation techniques. As error mitigation is applied, the fidelity witness
improves and error decreases. ``+MEM'' refers to measurement error mitigation.
``+PS'' refers to postselection of bitstrings based on parity. ``+Pur.'' refers to
purification of the measured correlation matrix. All circuits were executed with
dynamical decoupling pulses added.
(Bottom) The measured excitation energies of the first and second excitations above the
ground state, and their symmetric hole counterparts.
``Raw'' refers to raw data from circuits executed with dynamical decoupling pulses, and
``Mit.'' refers to the data after all error mitigation techniques are applied.
The signature of Majorana zero modes
is the zero-energy first excitation separated from the finite-energy second excitation
at $\mu=0$ and the robustness of this property to perturbations in $\mu$. This property
can be clearly seen in the two leftmost data points in each plot.
All error bars indicate a confidence interval of two standard deviations (95\% confidence)
as estimated from the sample covariance.
}
\label{fig:fidelity_witness_and_bdg_energy}
\end{figure*}

\begin{figure*}
\includegraphics[width=0.7\textwidth]{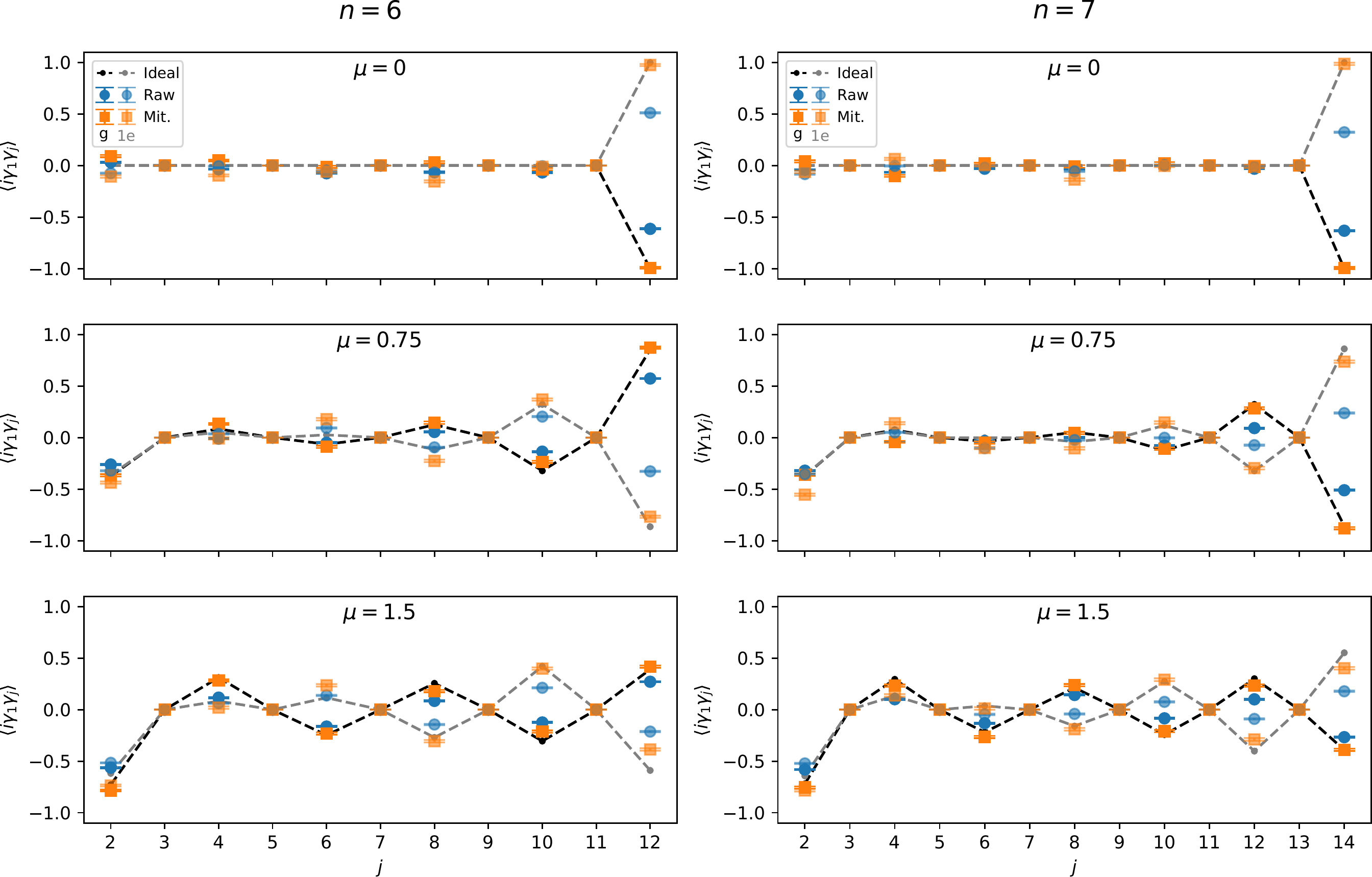}
\caption{Majorana site correlation. The Majorana site correlation at site $j$ is
defined as the operator $i \gamma_1 \gamma_j$ where $\gamma_j$ denotes the $j$-th Majorana operator.
The data for the 6-mode Kitaev chain is shown
on the left, and the data for the 7-mode Kitaev chain is shown on the right.
For each system size, we include plots for several different values of $\mu$,
as indicated on the plots. Each plot includes data for the ground state (ideal values
indicated by the dashed black line) and the first excited state (ideal values
indicated by the dashed gray line).
``Raw'' refers to raw data without error mitigation, and
``Mit.'' refers to the data after all error mitigation techniques are applied.
At $\mu = 0$ the correlation is only nonzero
for $j = 2n$, indicating an exclusive correlation between Majorana fermions
at the ends of the chain. As $\mu$ increases this exclusive correlation breaks down.
All error bars indicate a confidence interval of two standard deviations (95\% confidence)
as estimated from the sample covariance.
}
\label{fig:site_correlation}
\end{figure*}

\subsection{Circuits and observable measurement}

Since the Kitaev chain Hamiltonian is quadratic in the fermionic creation and annihilation
operators, its eigenstates are fermionic Gaussian states which can be prepared efficiently
using the algorithm given in Reference~\cite{jiang2018correlated}.
This algorithm has linear circuit depth
and requires only linear qubit connectivity. It assumes the Jordan-Wigner transform
is used to map fermionic operators to qubit operators.
Figure \ref{fig:circuit} displays an example circuit that shows
the general structure. Besides single-qubit Z rotations and X gates, the only other
type of gate present is the so-called Givens rotation gate, with matrix
\begin{align}
    G(\theta) = \begin{pmatrix}
                    1 & 0 & 0 & 0 \\
                    0 & \cos(\theta) & -\sin(\theta) & 0 \\
                    0 & \sin(\theta) & \cos(\theta) & 0 \\
                    0 & 0 & 0 & 1
                \end{pmatrix}.
\end{align}
On hardware for which CNOT is the native two-qubit interaction, this gate can
be implemented using the decomposition shown in Figure \ref{fig:xx_plus_yy},
which uses two CNOT gates in addition to single-qubit rotations.

Eigenstates of a quadratic Hamiltonian are prepared by effecting a basis change
that maps the fermionic creation operators $\set{a^\dagger_j}$
to a new set of creation operators $\set{b^\dagger_j}$ such that the Hamiltonian
takes the diagonal form
\begin{align}
    H = \sum_{j} \varepsilon_j b^\dagger_j b_j + \text{constant}
\end{align}
where $0 \leq \varepsilon_1 \leq \cdots \leq \varepsilon_n$.
The operators $\set{b^\dagger_j}$ also satisfy the fermionic anticommutation relations,
so they can be regarded as creation operators for fermionic excitations with
excitation energies given by $\set{\varepsilon_j}$.
The $\set{b^\dagger_j}$ are linear combinations of the original creation and annihilation operators:
\begin{align}
    \begin{pmatrix}
    b^\dagger_1 \\
    \vdots \\
    b^\dagger_n \\
    \end{pmatrix}
    = W
    \begin{pmatrix}
    a^\dagger_1 \\
    \vdots \\
    a^\dagger_n \\
    a_1 \\
    \vdots \\
    a_n
    \end{pmatrix}
\end{align}
where $W$ is an $n \times 2n$ matrix. The matrix $W$ can be efficiently computed from
the description of the Hamiltonian and is used to produce the quantum circuit that
prepares an eigenstate of the Hamiltonian.
In Appendix \ref{app:fermionic_gaussian_states}, we review the quantum algorithm
for preparing fermionic Gaussian states.

For each state prepared, all observables of interest can be
determined from the correlation matrix. The correlation matrix $\Gamma$
of a state $\rho$ is defined as the block matrix
\begin{align}
    \Gamma =
    \begin{pmatrix}
        T & S \\
        -S^* & I - T^T
    \end{pmatrix}
\end{align}
where
\begin{align}
    T_{jk} &= \tr \bracks*{a^\dagger_j a_k \rho} \\
    S_{jk} &= \tr \bracks*{a^\dagger_j a^\dagger_k \rho}
\end{align}
and $I$ is the identity matrix.

We measured the correlation matrix using a protocol similar to the one used in
Reference~\cite{arute2020hartreefock} to measure the
one-particle reduced density matrix (1-RDM),
which is the matrix $T$ in our notation.

The diagonal entries of the correlation matrix are occupation numbers
that can be measured straightforwardly in the computational basis.
To obtain the off-diagonal entries, we need to measure the operators
\begin{align}
    a_j^\dagger a_k + a_k^\dagger a_j \\
    -i(a_j^\dagger a_k - a_k^\dagger a_j) \\
    a_j^\dagger a_k^\dagger + a_k a_j \\
    -i(a_j^\dagger a_k^\dagger - a_k a_j)
\end{align}
Under the Jordan-Wigner transformation, the operators between
neighboring modes (i.e., $k = j+1$)
can be measured using a two-qubit basis change.
To measure the operators between all pairs of modes, we generate new
circuits by permuting the columns of $W$; each permutation
corresponds to a relabeling of the fermionic modes which cause new
pairs to become adjacent. This strategy allows us to measure the operators between all pairs of modes using the same circuit structure.
In total, $\ceil{n / 2} \times 8 + 1$ different circuits are required to measure
the full correlation matrix. In Appendix \ref{app:measurement}, we provide
a detailed description of our measurement protocol.

\subsection{Error mitigation}

We applied a number of error mitigation techniques to improve results.

\emph{Dynamical decoupling.}
Dynamical decoupling is a technique to reduce phase errors stemming from
time-correlated low-frequency noise
by applying refocusing pulses on idle qubits \cite{viola1999dynamical,ahmed2013robustness}.
There are many possible pulse sequences that could be applied.
We used the 4-pulse sequence
\begin{align}
   X_{\pi} \rightarrow Y_{\pi} \rightarrow X_{-\pi} \rightarrow Y_{-\pi}
\end{align}
where $X_{\pi}$ and $X_{-\pi}$ denote $X$ gates implemented using opposite sign
pulse amplitudes.
We used Qiskit \cite{qiskit} to schedule the dynamical decoupling pulses.
The software detects idle periods in the compiled circuit and in each
idle period inserts one pulse sequence, distributing the 4 pulses
with even temporal spacing.

\emph{Measurement error mitigation.}
The effect of measurement errors can be mitigated by treating it as a classical
noise channel and approximately inverting it \cite{bravyi2021mitigating}.
We used the software package \texttt{mthree} \cite{mthree, nation2021scalable} to
perform measurement error mitigation. The error mitigation procedure
converts raw bitstring counts into error-mitigated bitstring quasiprobabilities.
The error mitigation does not come for free; rather, there is increased uncertainty
in computed quantities.

\emph{Postselection of bitstrings.}
In Reference~\cite{arute2020hartreefock} which prepared the Hartree-Fock state, the ideal
final state had a well-defined particle number, so measured bitstrings with the incorrect
particle number could be discarded to improve results. In our case, the final state
does not have a well-defined particle number, but it does have a well-defined parity.
Therefore, we can still perform postselection on the bitstrings, discarding those
with the incorrect parity. A technical detail is that here we actually apply postselection not on the
bitstrings directly, but on the quasiprobability distribution over bitstrings
returned by the measurement error mitigation procedure.
In Table~\ref{tab:post_selection_fraction}, we show the quasiprobability mass
discarded by the postselection.

\emph{State purification.}
Reference~\cite{arute2020hartreefock} exploited the fact that the 1-RDM is idempotent
(it is equal to its square)
to perform purification of the measured 1-RDM. That is, due to experimental error,
the measured 1-RDM is not idempotent, and it was projected onto the space of
idempotent matrices using a procedure called
McWeeny purification~\cite{mcweeny1960some}.
In our case, it is the correlation matrix that is idempotent \cite{bach1994generalized}.
Therefore, McWeeny purification can also be applied here to purify the measured
correlation matrix. The purification is accomplished by repeating until
convergence the following numerical operation to update $\Gamma$ at iteration $k$:
\begin{align}
    \Gamma_{k + 1} = \Gamma_k^2 (3I - 2\Gamma_k).
\end{align}
\subsection{Experiments}

For our experiments, we set $t = -1$, $\Delta = 1$, and used 5 different values
for $\mu$ evenly spaced between 0 and 3.
For each choice of Hamiltonian parameters,
we prepared 6 eigenstates: the ground state and the first and second excited states, as well as the 3 corresponding eigenstates from the opposite end of the spectrum.
We executed our circuits on the
\textit{ibmq\_guadalupe} device accessed through the IBM Quantum service.
We used Qiskit~\cite{qiskit}
to compile the circuits into basis gates supported by the hardware.
The Givens rotation gates were compiled using a decomposition
similar to the one shown in Figure~\ref{fig:xx_plus_yy}, requiring
2 CNOT gates for each Givens rotation.
For each circuit we collected 100,000 measurement shots.

To execute each circuit, we needed to pick a line of qubits to use.
Because gate errors vary over the device, the choice of qubits can have a
significant impact on performance. To pick the qubits, we
used the software package \texttt{mapomatic} \cite{mapomatic}, which attempts to
minimize the expected error in executing the circuit using a subgraph isomorphism 
algorithm scored on gate errors reported for the device.

Reference~\cite{gluza2018witness} shows how to compute a fidelity witness
for experiments that prepare fermionic Gaussian states.
The fidelity witness gives a lower bound
on the fidelity of the experimentally measured state with the ideal state
and can be easily computed from the correlation matrix;
see Appendix \ref{app:fidelity_witness} for a review of this result.
Figure~\ref{fig:fidelity_witness_and_bdg_energy} (top panel)
shows the fidelity witness and average error in energy.

Figure~\ref{fig:fidelity_witness_and_bdg_energy} (bottom panel) shows the
measured excitation energies of the first and second excitations above the
ground state, and their symmetric hole counterparts. We show both the
values obtained from the raw data and those obtained after applying
error mitigation.

Figure~\ref{fig:site_correlation} shows the measured expectation value of the
Majorana site correlation $i \gamma_1 \gamma_j$. Again, both raw and error-mitigated
data are displayed.

\begin{table}
\begin{tabular}{|c|c|c|}
    \hline
    System size & Vacuum & Occupied \\ \hline \hline
    6 qubits & 0.283 & 0.348 \\ \hline
    7 qubits & 0.344 & 0.398 \\ \hline
\end{tabular}
\caption{Quasiprobability mass discarded by postselection on bitstring parity.
The ``Vacuum'' column indicates circuits where the initial state is the all
zeros state; due to an optimization, these circuits had fewer gates than
circuits with an ``occupied'' initial state, and hence a lower expected fraction
of bitstrings with the incorrect parity.
}
\label{tab:post_selection_fraction}
\end{table}

\begin{figure}
\includegraphics[width=\linewidth]{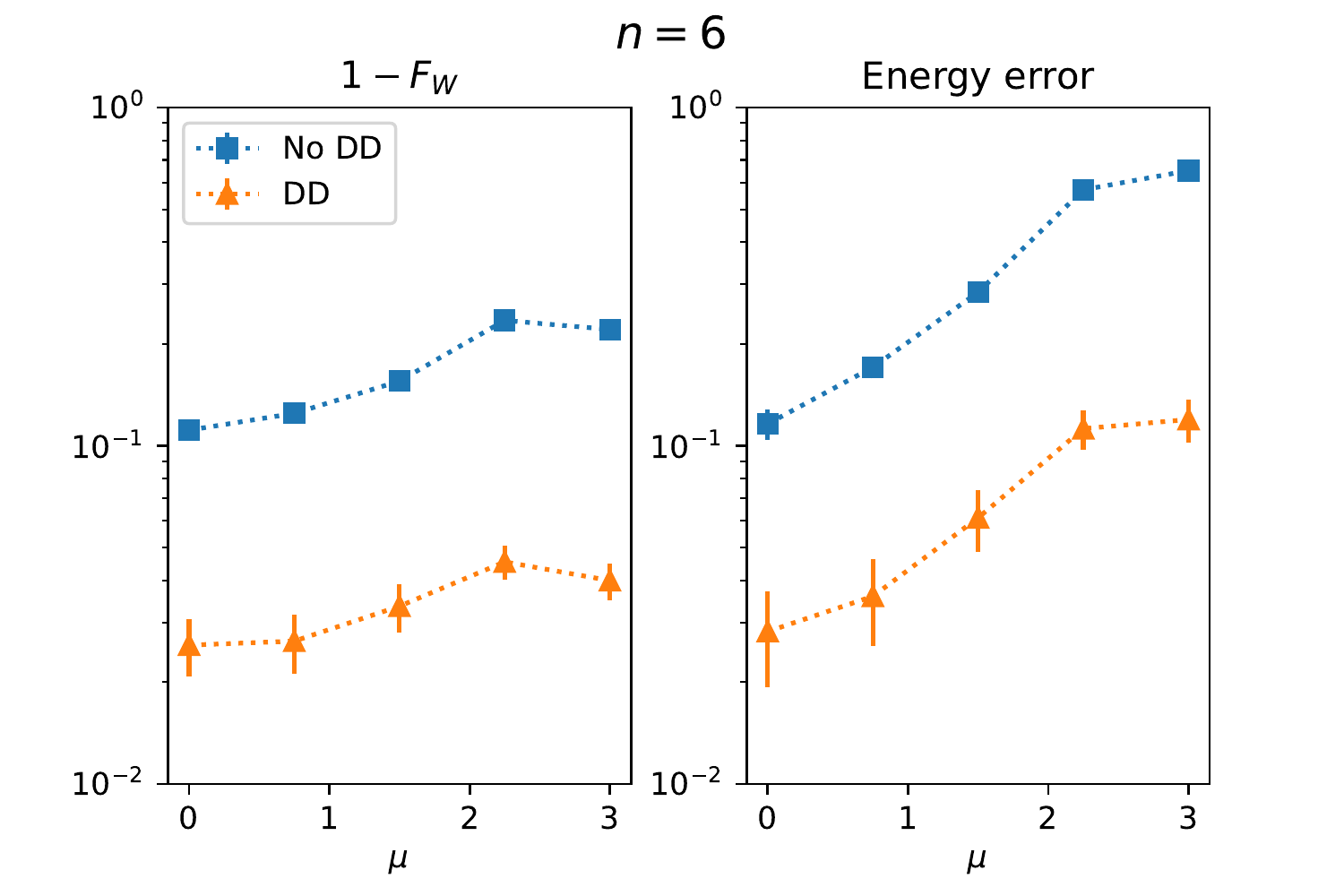}
\caption{Effectiveness of dynamical decoupling. This figure shows the
fidelity witness and the energy error for experiments
without dynamical decoupling (No DD)
and with dynamical decoupling (DD) on the Kitaev chain of length 6.}
\label{fig:fidelity_witness_dd_compare}
\end{figure}

\section{Discussion}

We created Majorana zero modes on a noisy superconducting qubit processor by preparing
eigenstates of the Kitaev chain Hamiltonian. The largest chain that we simulated
used 7 qubits. Simulating this chain required executing circuits containing
dozens of two-qubit gates, yet our experimental results
closely matched theory despite their implementation on noisy hardware.
We measured zero-energy excitations at $\mu = 0$ which
were robust to small perturbations in $\mu$. We also observed
the exclusive correlation between Majorana fermions at the ends of the chain at $\mu=0$ and increasing correlations with interior sites with increasing $\mu$.
In the limit of infinite chain length, the value of $\mu = 2t$ separates
two topological phases, but with a finite length chain this value is smaller; at $\mu = 1.5t$ in our experiments the correlation
between the first site and interior sites is as strong as that between the ends of the chain. At small chemical potential, the wavefunction for a finite-length chain decays exponentially as $\exp{\left(-(j-1)/\xi\right)}$, where $j$ is the site and $\xi$ is the characteristic MZM decay length \cite{leumer2020exact}

\begin{equation}
\xi=2 \begin{cases}
      \left| \ln{\left(\frac{t-\Delta}{t+\Delta}\right)} \right|^{-1}& |t|\ge |\Delta|, \\
      \left| \ln{\left(\frac{\Delta-t}{\Delta+t}\right)} \right|^{-1} & |\Delta|\ge |t|. 
    \end{cases}  
\end{equation}

The quality of our results was made possible by our
utilization of error mitigation techniques.
We applied dynamical decoupling pulses to idle qubits to mitigate
dephasing errors. To see whether these pulses actually improved results,
we ran experiments with and without dynamical decoupling (while still applying
the rest of the error mitigation techniques).
Figure \ref{fig:fidelity_witness_dd_compare} shows the fidelity witness
and energy error for these experiments on the 6-mode Kitaev chain.
Applying dynamical decoupling pulses does indeed yield a significant improvement, which suggests that low frequency noise on idle qubits is a major source of dephasing error.
We also applied measurement error mitigation;
this is a straightforward and widely applicable technique that is becoming
common practice.

The other techniques we applied were
physically-motivated postselection of measured bitstrings and state purification.
These techniques have previously been used in the preparation of
Slater determinants \cite{arute2020hartreefock}. Here, we showed how to
extend these techniques to the general case of fermionic Gaussian states so they can
be used for the preparation of arbitrary eigenstates of
quadratic Hamiltonians such as the Kitaev chain Hamiltonian.
These techniques were highly effective at improving the fidelity and lowering
the error of our simulations.

The combined application of all of these error mitigation techniques
enabled us to simulate systems of up to 7 qubits, whereas a
previous experiment preparing Majorana zero modes used only 3 qubits.
We note that References~\cite{harle2022observing,mi2022majorana}
performed larger experiments (up to 21 and 47 qubits, respectively)
with Majorana modes on noisy quantum processors
by periodically driving the qubits with a Floquet unitary,
implementing time-varying Hamiltonians in both non-interacting and
interacting regimes. Those works do not directly prepare eigenstates, but rather
extract experimental signatures of the Majorana modes via spectroscopic measurements.
This contrasts with our work, which focuses on the exact preparation
and measurement of eigenstates of the static Kitaev model.

Although in this work we deal with the non-interacting Kitaev chain model,
a modified version of the model with an added interaction term has received
much attention
\cite{fidkowski2010effects,fidkowski2011topological,gangadharaiah2011majorana,rahmani2015emergent,miao2017exact,Stenger2022}. The interacting model does not have an analytic solution in the
general case.
Its eigenstates are not fermionic Gaussian states, so the algorithm we used
in this work cannot be applied directly to its solution.
However, our work does open the path to the study of the interacting Kitaev chain
on quantum computers. Depending on the strength of the interaction, the ground state
of the model may have a non-negligible overlap with the ground state of a non-interacting
model. Thus, the ground state of the non-interacting model, prepared using the methods in
this work, may be used as the initial state for a
variational \cite{peruzzo2014variational,cerezo2021variational}
or adiabatic \cite{farhi2000adiabatic,aharonov2007adiabatic} algorithm
for preparing the ground state of the interacting model.

Our results build on previous experiments suggesting that error mitigation
will be crucial to achieving practical applications on noisy near-term
quantum computers,
and they contribute to the growing library of experiments that can serve
as device benchmarks as we work towards those practical applications.

\subsection*{Acknowledgments}

The authors acknowledge the use of IBM Quantum Services for this work.
This experiment was implemented and executed using
Qiskit~\cite{qiskit}.
The circuit diagrams in this paper were created using
$\langle\mathsf{q}|\mathsf{pic}\rangle$~\cite{qpic}.
The data plots were
created using Matplotlib~\cite{matplotlib}.

\subsection*{Data availability}

The experimental data for this work is available at \url{https://doi.org/10.5281/zenodo.6603265}.

\subsection*{Code availability}

A snapshot of the source code is included at \url{https://doi.org/10.5281/zenodo.6603265}.
An up-to-date version is maintained at
\url{https://github.com/qiskit-research/qiskit-research}.

\bibliography{main}

\subsection*{Author contributions}

N.T.B. and O.T.L. conceived the project.
K.J.S. developed the error mitigation techniques
and collected and analyzed the experimental data.
K.J.S. and N.T.B. wrote the software for the experiment.
K.J.S. and M.J.R. wrote the manuscript with the assistance of N.T.B. and O.T.L.
All authors contributed to discussions that shaped the project.

\subsection*{Competing interests}

The authors declare no competing interests.

\appendix
% \setcounter{table}{0}
% \renewcommand{\thetable}{S\arabic{table}}%
% \setcounter{figure}{0}
% \renewcommand{\thefigure}{S\arabic{figure}}%

% \newpage

\section{Fermionic Gaussian states}
\label{app:fermionic_gaussian_states}

In this section, we provide background material on fermionic Gaussian states
and review the quantum algorithm given in Reference \cite{jiang2018correlated} for preparing them.

Fermionic Gaussian states can be defined in several equivalent ways. One definition is that
fermionic Gaussian states are eigenstates of quadratic fermionic Hamiltonians.
The general form of a quadratic fermionic Hamiltonian can be written as
\begin{align}
  H = \sum_{p, q = 1}^n M_{pq} a^\dagger_p a_q +
  \frac12 \sum_{p, q = 1}^n (\Delta_{pq} a^\dagger_p a^\dagger_q - \Delta_{pq}^* a_p a_q),
  \label{eq:quad_ham}
\end{align}
where $M$ and $\Delta$ are $n \times n$ matrices.
The operators $\set{a_p}_{p=1}^n$ are fermionic annihilation operators which
satisfy the fermionic anticommutation relations
\begin{align}
a_p a_q + a_q a_p &= 0, \label{eq:car1} \\
a_p a^\dagger_q + a^\dagger_q a_p &= \delta_{pq}. \label{eq:car2}
\end{align}
The adjoint $a^\dagger_p$ of an annihilation operator $a_p$ is called a creation operator.
Since $H$ is Hermitian,
we must have $M = M^\dagger$ and $\Delta = -\Delta^T$.
Any quadratic Hamiltonian can be rewritten in the following form:
\begin{align}
  H = \sum_{p = 1}^n \eps_p b^\dagger_p b_p + \text{constant},
  \label{eq:quad_ham_diag}
\end{align}
where the $\eps_p$ are non-negative real numbers. The $\set{b_p}$ are a new set of
fermionic annihilation operators that also satisfy the anticommutation relations
\eqref{eq:car1} and \eqref{eq:car2}. They are linear combinations of
the original creation and annihilation operators:
\begin{align}
  \begin{pmatrix}
    b^\dagger_1 \\
    \vdots \\
    b^\dagger_n \\
    b_1 \\
    \vdots \\
    b_n
  \end{pmatrix}
  = W
  \begin{pmatrix}
    a^\dagger_1 \\
    \vdots \\
    a^\dagger_n \\
    a_1 \\
    \vdots \\
    a_n
  \end{pmatrix},
  \label{eq:W}
\end{align}
where $W$ is a $2n \times 2n$ matrix. The matrix $W$ is unitary and has the block form
\begin{align}
  W =
  \begin{pmatrix}
    W^*_1 & W^*_2 \\
    W_2 & W_1
  \end{pmatrix},
  \label{eq:bog_transform}
\end{align}
where the fermionic anticommutation relations imply that
\begin{align}
  W_1 W_2^T + W_2 W_1^T &= 0, \label{eq:W_cond_1}\\
  W_1 W_1^\dagger + W_2 W_2^\dagger &= I. \label{eq:W_cond_2}
\end{align}
The matrix $W$ can be efficiently computed from $M$ and $\Delta$ using
a Schur decomposition; we refer the reader to Appendix A of Reference \cite{jiang2018correlated}
for a detailed description of how to perform this computation.
Source code for performing this computation is available in both OpenFermion \cite{openfermion}
and Qiskit Nature \cite{qiskit}.

A fermionic Gaussian state can be prepared as an eigenstate of the Hamiltonian (\ref{eq:quad_ham})
by effecting a unitary transformation $\calW$ such that
\begin{align}
  \calW a_p \calW^\dagger = b_p, \qquad p = 1, \ldots, n.
\end{align}
Up to a global phase, this unitary is determined by the matrix $W$.
The ground state of $H$ is
\begin{align}
  \ket{\Psi_0} = \calW \vac,
  \label{eq:gauss_ground_state}
\end{align}
and the other eigenstates are of the form
\begin{align}
  \parens*{b^\dagger_1}^{i_1} \cdots \parens*{b^\dagger_n}^{i_n} \ket{\Psi_0},
  \label{eq:gauss_state}
\end{align}
where $i_p \in \set{0, 1}$ for $p = 1, \ldots, n$. Therefore, a fermionic
Gaussian state is prepared by applying $\calW$ to a state of the form
\begin{align}
\label{eq:comp_basis_vec}
\parens*{a^\dagger_1}^{i_1} \cdots \parens*{a^\dagger_n}^{i_n} \vac,
\end{align}
a computational basis state under the Jordan-Wigner transformation.

To implement the unitary, we used the algorithm described in Reference \cite{jiang2018correlated}.
The algorithm starts with the matrix $W$
as input. Actually, due to the redundancy in $W$ only the lower half $W_L$ is used:
\begin{align}
  W_L = (W_2 \quad W_1).
  \label{eq:W_L}
\end{align}
The algorithm works by finding a decomposition of $W_L$
\begin{align}
  V W_L U^\dagger = (0 \quad I),
  \label{eq:W_canonical}
\end{align}
where $I$ is the identity matrix, $V$ is an $n \times n$ unitary matrix,
and $U$ is further decomposed into elementary matrix operations as
\begin{align}
  U = B G_{n_G} \cdots B G_{3} G_{2} B G_{1} B.
\end{align}
Here, each $G_i$ is a matrix of the form
\begin{align}\label{eq:G}
G(\theta, \ph) = \left(
\begin{array}{cc|cc}
  \cos\theta  & -e^{i\varphi} \sin \theta & 0& 0\\
  \sin \theta &  e^{i\varphi} \cos \theta   & 0& 0\\
  \hline
  0& 0 &\cos\theta  & -e^{-i\varphi} \sin \theta \\
  0& 0 &\sin \theta &  e^{-i\varphi} \cos \theta  
\end{array}
\right),
\end{align}
and
\begin{align}\label{eq:F}
B = B^\dagger =
\begin{pmatrix}
  I-e_n e_n^T & e_n e_n^T \\[4pt]
  e_n e_n^T & I-e_n e_n^T
\end{pmatrix},
\end{align}
where $e_n = (0, \ldots, 0, 1)^T$ is a vector of length $n$.
The decomposition (\ref{eq:W_canonical}) corresponds to a decomposition
\begin{align}
  \calW = \calB \calG_1 \calB \calG_2 \calG_3 \calB \cdots \calG_{n_G} \calB \cdot \calV
  = \calU \cdot \calV
  \label{eq:ferm_gauss_decomp}
\end{align}
of the desired unitary in terms of elementary gates. Here
each $\calG_i$ is a complex Givens rotation gate of the form
\begin{align}
  \calG(\theta, \ph) = \exp\bracks*{i \ph a_q^\dagger a_q}
  \exp\bracks*{\theta (a_p^\dagger a_q - a_q^\dagger a_p)}
  \label{eq:givens_ferm}
\end{align}
acting on neighboring modes,
and $\calB$ is the particle-hole transformation on the last fermionic mode,
\begin{align}
  \calB a_n \calB^\dagger &= a_n^\dagger, \\
  \calB a_p \calB^\dagger &= a_p \qquad \text{for } p = 1, \ldots, n-1.
  \label{eqref:bogo}
\end{align}
Under the Jordan-Wigner transformation, the Givens rotation
between neighboring modes is a two-qubit gate with matrix
\begin{align}
    G(\theta, \ph) = \begin{pmatrix}
                    1 & 0 & 0 & 0 \\
                    0 & e^{i\ph} \cos(\theta) & -e^{i\ph} \sin(\theta) & 0 \\
                    0 & \sin(\theta) & \cos(\theta) & 0 \\
                    0 & 0 & 0 & e^{i\ph}
                \end{pmatrix},
\end{align}
and the particle-hole
transformation is a single-qubit $X$ gate on the last qubit.
The unitary $\calV$ can also be decomposed into Givens rotations,
but this is not necessary when the initial state of the circuit is
the all zeros state, i.e., when the ground state is being prepared.
The total number of Givens rotations used by the algorithm is
\begin{align}
    \frac{n (n - 1)}{2} + \eta (n - \eta)
\end{align}
where $\eta$ is the number of pseudoparticles.
The circuit depth is at most $3n - 2$.

In summary, this algorithm yields a quantum circuit for preparing the
fermionic Gaussian state under the Jordan-Wigner transformation with
linear depth and which uses only linear connectivity.
Source code for this algorithm is available in both
OpenFermion~\cite{openfermion} and Qiskit Nature~\cite{qiskit}.

\section{Measurement of the correlation matrix}

\label{app:measurement}

In this section, we provide details on our strategy for measuring the
correlation matrix of a fermionic Gaussian state. Our strategy is similar to the one
used in Reference~\cite{arute2020hartreefock} to measure the
one-particle reduced density matrix (1-RDM).

The correlation matrix $\Gamma$
of a state $\rho$ is defined as the block matrix
\begin{align}
    \Gamma =
    \begin{pmatrix}
        T & S \\
        -S^* & I - T^T
    \end{pmatrix}
\end{align}
where
\begin{align}
    T_{jk} &= \tr \bracks*{a^\dagger_j a_k \rho} \\
    S_{jk} &= \tr \bracks*{a^\dagger_j a^\dagger_k \rho}
\end{align}
and $I$ is the identity matrix.

The diagonal entries of the correlation matrix are occupation numbers
that can be measured straightforwardly in the computational basis.
To obtain the off-diagonal entries, we need to measure the operators
\begin{align}
    a_j^\dagger a_k + a_k^\dagger a_j &\mapsto \frac{X_j X_k + Y_j Y_k}{2} Z_{j+1} \cdots Z_{k - 1} \\
    -i(a_j^\dagger a_k - a_k^\dagger a_j) &\mapsto \frac{X_j Y_k - Y_j X_k}{2} Z_{j+1} \cdots Z_{k - 1} \\
    a_j^\dagger a_k^\dagger + a_k a_j &\mapsto \frac{X_j X_k - Y_j Y_k}{2} Z_{j+1} \cdots Z_{k - 1} \\
    -i(a_j^\dagger a_k^\dagger - a_k a_j) &\mapsto \frac{-X_j Y_k - Y_j X_k}{2} Z_{j+1} \cdots Z_{k - 1}
\end{align}
where $j < k$ and we have shown how the operators map under the Jordan-Wigner transformation.
For the case of neighboring modes, i.e., $k = j+1$, we have
\begin{align}
    a_j^\dagger a_k + a_k^\dagger a_j &\mapsto \frac{X_j X_k + Y_j Y_k}{2} \label{eq:meas_ops_1} \\
    -i(a_j^\dagger a_k - a_k^\dagger a_j) &\mapsto \frac{X_j Y_k - Y_j X_k}{2} \label{eq:meas_ops_2} \\
    a_j^\dagger a_k^\dagger + a_k a_j &\mapsto \frac{X_j X_k - Y_j Y_k}{2} \label{eq:meas_ops_3} \\
    -i(a_j^\dagger a_k^\dagger - a_k a_j) &\mapsto \frac{-X_j Y_k - Y_j X_k}{2}. \label{eq:meas_ops_4}
\end{align}

If the correlation matrix is real, then only the first and third operators need
to be measured. A sufficient condition for the correlation matrix to be real is that
the circuit used to prepare the state contains only gates with real-valued
matrices, up to global phase. In our experiment, we checked for this condition
by directly inspecting the gates in the circuit.

Between neighboring modes, these operators can be measured by diagonalizing them with a
parity-preserving two-qubit gate. For example, the third operator is diagonalized
by the gate with matrix
\begin{align}
    \begin{pmatrix}
        \frac{1}{\sqrt{2}} & 0 & 0 & \frac{1}{\sqrt{2}} \\
        0 & 1 & 0 & 0 \\
        0 & 0 & 1 & 0 \\
        -\frac{1}{\sqrt{2}} & 0 & 0 & \frac{1}{\sqrt{2}}
    \end{pmatrix}.
\end{align}
Using parity-preserving gates for measurement enables the error mitigation
technique of postselection on bitstring parity, described below.
These gates can be implemented on hardware similarly to the Givens
rotation gates by decomposing them into two CNOT gates plus single-qubit
rotations.

To measure the operators between non-neighboring modes, there are several possible approaches.
One could measure them directly in the Pauli basis using single-qubit rotations to effect the
basis changes. This approach has the drawback that the single-qubit rotations do not preserve the parity,
so we would not be able to perform bitstring postselection on parity. Alternatively, one could
perform fermionic swap gates to swap the modes until they are neighboring. However, this
approach involves adding additional gates to the circuits, which increases the errors due to
execution on a noisy device.

Here, we take a different approach, which was described in
Reference~\cite{arute2020hartreefock} for the case of measuring the 1-RDM
of a Slater determinant.
Recall that the fermionic Gaussian state is prepared
by effecting a unitary transformation $\calW$ such that
\begin{align}
  \calW a_p \calW^\dagger = b_p, \qquad p = 1, \ldots, n.
\end{align}
The $\set{b^\dagger_j}$ are linear combinations of the original creation and annihilation operators:
\begin{align}
    \begin{pmatrix}
    b^\dagger_1 \\
    \vdots \\
    b^\dagger_n \\
    \end{pmatrix}
    = W
    \begin{pmatrix}
    a^\dagger_1 \\
    \vdots \\
    a^\dagger_n \\
    a_1 \\
    \vdots \\
    a_n
    \end{pmatrix}
\end{align}
where $W$ is an $n \times 2n$ matrix.
The matrix $W$ can be efficiently computed from
the description of the Hamiltonian and is used to produce the quantum circuit that
prepares an eigenstate of the Hamiltonian, as described in Appendix \ref{app:fermionic_gaussian_states}.
Note that the columns of $W$ are indexed by the operators $\set{a^\dagger_j}$
and $\set{a_j}$. However, the ordering of the operators $\set{a^\dagger_j}$ is arbitrary, so we can reorder them without changing the
definition of the $\set{b^\dagger_j}$ and the target fermionic Gaussian state.
A reordering of the operators $\set{a^\dagger_j}$ corresponds
to a permutation of the first $n$ columns of $W$. We want to use the same order
for both the $\set{a^\dagger_j}$ and the $\set{a_j}$, so we will always apply the
same permutation to the first
$n$ columns of $W$ and the last $n$ columns.
Each permutation causes different pairs of
modes to be mapped to neighboring qubits. This procedure enables the measurement of operators
between all pairs of modes without the need to add fermionic swap gates to the circuits.
By generating the permutations according to a ``parallel bubble sort'' pattern,
we can minimize the total number of permutations needed.

As an example, consider a system of 6 fermionic modes, initially labeled as
\begin{align}
    (0, 1, 2, 3, 4, 5)
\end{align}
In this configuration, we can measure the operators between 0--1, 1--2, etc.
The parallel bubble sort pattern performs a sequence of swaps starting on even indices,
followed by a sequence of swaps on odd indices:
\begin{align}
    (0, 1, 2, 3, 4, 5) \rightarrow (1, 0, 3, 2, 5, 4) \rightarrow (1, 3, 0, 5, 2, 4).
\end{align}
Now, we can measure the operators between 1--3, 0--5, etc. Repeating this
procedure one more time yields the configuration (3, 5, 1, 4, 0, 2), allowing
the rest of the operators to be measured. In total, $\ceil{n / 2}$ permutations
are required, including the identity permutation corresponding to the initial configuration. For each permutation, four
different basis changes are required
to measure the operators (\ref{eq:meas_ops_1}-\ref{eq:meas_ops_4}), and
each basis change gives rise to two circuits, one to measure pairs of qubits
starting on even indices, and one for odd indices.
Together with the circuit for measuring the diagonal entries,
measuring the correlation matrix of a state requires
$\ceil{n / 2} \times 8 + 1$ circuits in total. In our experiments
the correlation matrix was real, so only two of the basis changes
were required and the total number of circuits was
$\ceil{n / 2} \times 4 + 1$.

\begin{figure}
\includegraphics[width=\linewidth]{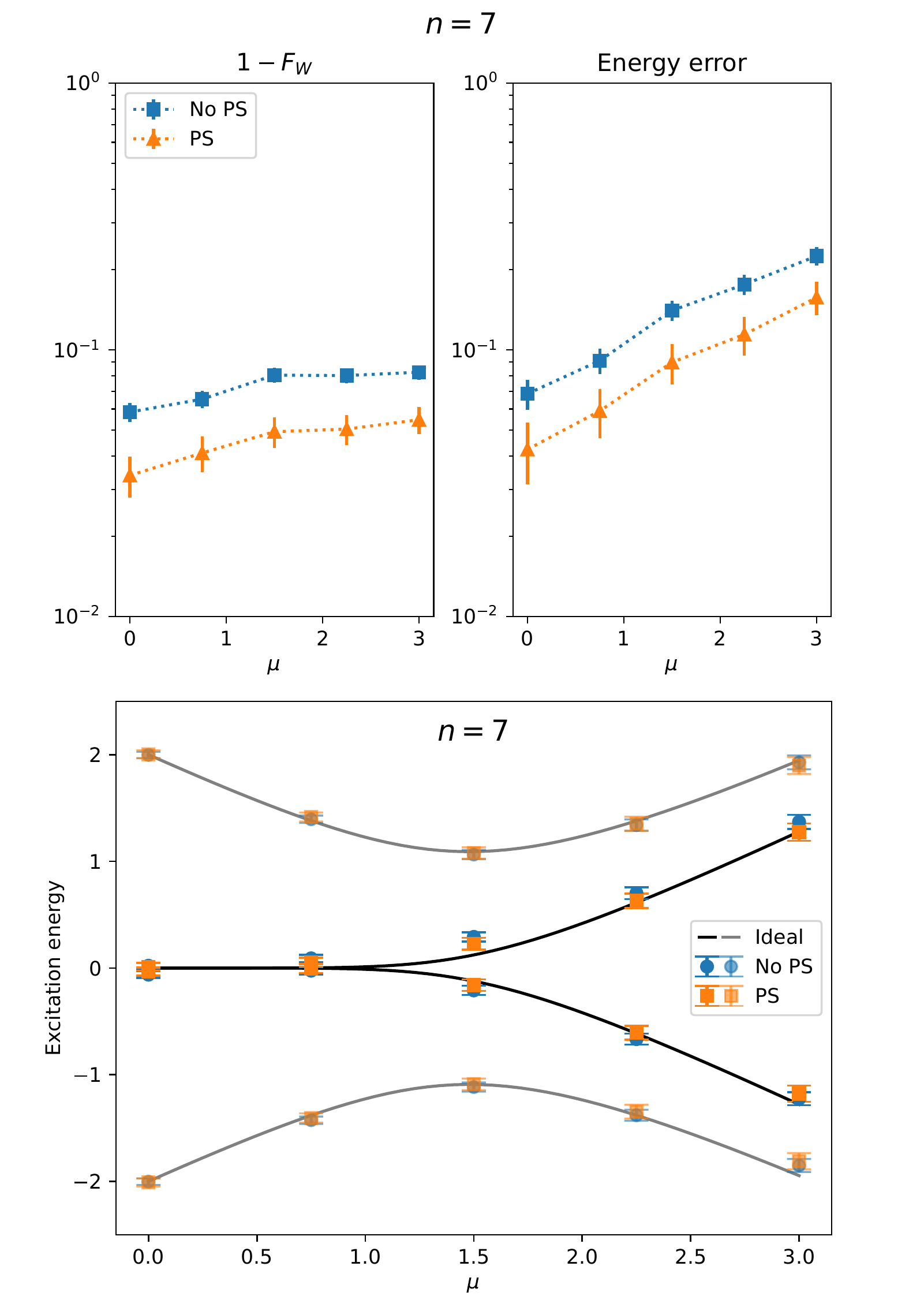}
\caption{Effectiveness of post-selection. Data is presented for the
Kitaev chain of length 7. (Top) Fidelity witness and
energy error for the data analyzed
without postselection (No PS)
and with postselection (PS). (Bottom) Measured excitation energies
with and without postselection.}
\label{fig:ps_compare}
\end{figure}

\section{Fidelity witness}
\label{app:fidelity_witness}

Reference \cite{gluza2018witness} shows how to obtain a fidelity witness for experiments
preparing fermionic Gaussian states. In this section we review this result and describe how
we computed the fidelity witness in our experiment.

Consider an experiment that aims to prepare a known pure fermionic Gaussian target state $\rho_t$.
Let $\rho_p$ denote the imperfect state that is actually prepared in the experiment.
The closeness between the two states is measured by the fidelity
\begin{align}
    F(\rho_t, \rho_p) = \parens*{\tr \bracks*{\parens*{\sqrt{\rho_t} \rho_p \sqrt{\rho_t}}^\frac12}}^2.
\end{align}
Since $\rho_t$ is pure, this equation simplifies to
\begin{align}
    F(\rho_t, \rho_p) = \tr \bracks*{\rho_t \rho_p}.
\end{align}
A fidelity witness is an observable $W$ for which the value $F_W(\rho_p) = \tr \bracks*{W \rho_p}$ gives a lower bound
on the fidelity between $\rho_t$ and $\rho_p$.
Reference \cite{gluza2018witness} describes such a witness and gives an expression for the fidelity lower bound in terms
of the covariance matrices $M_t$ and $M_p$ of $\rho_t$ and $\rho_p$:
\begin{align}
    F_W(\rho_p) = 1 + \frac14 \tr \bracks{\parens*{M_p - M_t}^T M_t}
\end{align}
The covariance matrix $M$ of a state $\rho$ has entries
\begin{align}
    M_{jk} = \frac{i}{2} \tr \bracks*{\parens*{\gamma_j \gamma_k - \gamma_k \gamma_j} \rho}
\end{align}
where the $\gamma_j$ are Majorana fermion operators.
To relate the covariance matrix to the correlation matrix it is
convenient to use the following alternative indexing convention for
the Majorana operators:
\begin{align}
    \gamma_{j} = a_j + a^\dagger_j, \quad \gamma_{j + n} = -i \parens*{a_j - a^\dagger_j}.
\end{align}
Under this convention, the covariance matrix is related to the correlation matrix $\Gamma$ by the identity
\begin{align}
    M = i \Omega \parens*{2 \Gamma - I} \Omega^\dagger
\end{align}
where $\Omega$ is the block matrix
\begin{align}
    \Omega = \frac{1}{\sqrt{2}}
    \begin{pmatrix}
        I & I \\
        iI & -iI
    \end{pmatrix}.
\end{align}
Using this relation, the expression for the fidelity lower bound can be written in terms of the correlation
matrices $\Gamma_t$ and $\Gamma_p$ of $\rho_t$ and $\rho_p$:
\begin{align}
    \label{eq:fidelity_witness}
    F_W(\rho_p) = 1 - \tr \bracks*{\parens*{\Gamma_t - \Gamma_p} \parens*{\Gamma_t - \frac{I}{2}}}.
\end{align}

While Reference \cite{gluza2018witness} describes an efficient protocol for measuring the fidelity lower bound
without needing to measure the entire covariance or correlation matrix,
in our experiment we measured the entire correlation matrix and directly used Equation~(\ref{eq:fidelity_witness})
to compute the fidelity lower bound.

\section{Effect of postselection}

In this section, we provide additional data demonstrating the effect of
the error mitigation technique of postselection on bitstring parity.
Figure \ref{fig:ps_compare} shows the fidelity witness, energy error,
and excitation energies for the data from the 7-qubit experiment,
analyzed with and without postselection. All the other error mitigation
techniques are still applied. The data shows that while postselection
does give a noticeable improvement to fidelity and energy precision,
the energy curves are still reproduced
quite well without it, including the measurement of the zero-energy
excitation states.

\end{document}